\documentclass[10pt]{article}
\usepackage{moriond,epsfig}

\def\Journal#1#2#3#4{{#1} {\bf #2}, #3 (#4)}

\def\NPBP{{\em Nucl. Phys.} B (proc. supp.)}
\def\PLB{{\em Phys. Lett.}  B}
\def\PRL{\em Phys. Rev. Lett.}
\def\PRD{{\em Phys. Rev.} D}

\def\kevee{keV$_{ee}$ } 

\begin{document}
\vspace*{4cm}

\title{THE EDELWEISS EXPERIMENT AND DARK MATTER DIRECT DETECTION}

\author{ V. SANGLARD \it{for the EDELWEISS collaboration~\cite{edelweiss2}} }

\address{Institut de Physique Nucl\'eaire de Lyon,\\IN2P3-CNRS,Universit\'e
Claude Bernard Lyon I,\\4, rue Enrico Fermi,
69622 Villeurbanne, France}

\maketitle\abstracts{
This mini-review first introduces the motivations for Dark Matter Searches.
The experimental aspect of the direct detection of Weakly 
Interactive Massive Particles (WIMPs) is described, detailing its principle and presenting 
some experiments with their recent results. The EDELWEISS experiment  
and its results are discussed in more details before concluding with the future 
direct detection experiments.}

\section{Introduction}
\label{sec:intro}

\subsection{Why Dark Matter ?}
\label{subsec:why}

One of the most important questions in physics today is the problem of the Universe mass.
 It is one of the topics that motivated the association of astrophysics and 
particle physics in a new field : astroparticle physics.\\
In 1933 F. Zwicky first proposed a dark component of Universe~\cite{zwic}. By
measuring the velocity dispersion of galaxies in the Coma cluster, he highlighted the
fact that the Universe must contain something else in addition to visible matter : 
another form of matter which does not emit nor absorb radiation. In the seventies, this 
study was systematically extended to spiral galaxies by several 
teams. It was observed that the rotation velocity of galaxy arms remained constant as a 
function of distance from the center, even far beyond the luminous disc~\cite{halo}. This 
suggested that all galaxies are surrounded by a Dark Matter halo.\\
The hypothesis of Dark Matter has also gained momentum in the field of cosmology, as experiments 
are now determining with an increasing precision the cosmological parameters of the
Universe.\\
Recently new interesting results reached us from the balloon-borne experiment Archeops~\cite{arch} 
and the WMAP satellite~\cite{WMAP}. These two experiments
observe the CMB (Cosmic Microwave Background) fluctuations. Their results support a flat 
Universe with an energy density~\cite{arch} $\Omega = 1.00 \pm 0.03$ (normalized to 
the critical density). In addition, a gravitationaly self-repulsive "dark energy" accelerating the expansion
of the Universe and accounting 
for $\sim$ 70$\%$ of the total energy density, leaving only 30$\%$ for the matter one, has been evidenced, 
in agreement with the results of 
experiments studying distant type Ia SuperNovae~\cite{perlm}. Their data combined with other CMB
results constrain the baryon contents of the Universe~\cite{arch,WMAP} to a 
small fraction of matter $\Omega_b = 0.044 \pm 0.003$, but compatible with the primordial nucleosynthesis 
scenario. Thus standard cosmology supports an important presence of non-baryonic Dark Matter, with a 
density $\Omega_{DM} \approx$ 0.3. Therefore most of the matter in the Universe is dark and only a 
small part is baryonic. 

\subsection{Dark Matter candidates}
\label{subsec:cand}

Despite the growing acceptance for the existence of non-baryonic Dark Matter, its exact nature remains 
mysterious for a great part. It is worthwhile to go to particle physics to find well motivated candidates 
for Dark Matter.\\
Non-baryonic Dark Matter has two components : HDM (Hot Dark Matter) which was relativistic at the time when 
radiation decoupled from the matter and CDM (Cold Dark Matter), non relativistic at that time. The neutrino 
is the main candidate for HDM, but the experiments dedicated to the CMB fluctuation observations
strongly constrain the neutrino density~\cite{WMAP} $\Omega_{\nu} h^2 <$ 0.0076 (95 $\%$ C.L.), with 
$h$ = 0.71 $\pm$ 0.04. This paper will concentrate on CDM and in particular on the candidate with the best 
physics supports : the 
WIMP (Weakly Interactive Massive Particle) \footnote{A second candidate is the axion, a Goldstone boson which 
could permit to solve the CP violation problem in the strong interactions (for more details, see 
\cite{axion}).}. The WIMP has no electric charge and is a stable thermal relic from the Big Bang era, now
trapped in the gravitational potential of galaxies and clusters of galaxies. 
On the other side, CMB experiments suggest a CDM density~\cite{WMAP2} $\Omega_{CDM} \approx 0.22$; that 
confirms their importance in the galactic halo.\\
In the MSSM (Minimal Supersymmetric Standard Model) framework with the R-parity conservation, the LSP (Lightest 
Supersymmetric Particle) is a stable particle with a relic abundance. The most likely LSP is the neutralino, defined as a linear combination of the supersymmetric 
partners of the photon, Z and Higgs bosons. The neutralino mass ranges between 45 GeV/c$^2$ (given by LEP 
\cite{masse}) and a few TeV/c$^2$ ~\cite{mmm}. The neutralino is expected to be detected by an accelerator 
experiment like the LHC~\cite{LHC} (Large Hadron Collider).\\
A great number of experiments all over the world are aiming at discovering Dark Matter in the form of WIMPs, and in 
particular as a neutralino or another particle with a similar behavior.  

\section{WIMP search}
\label{subsec:search}

There are two different methods to observe WIMPs : direct and indirect detection. Indirect detection is 
the observation of the products of WIMP annihilation in cosmic rays. This technique is 
discussed elsewhere~\cite{indirect}. This paper is devoted to direct detection. We will describe its 
principles and present some of the leading experiments.\\
For the sake of comparing different experiments, it is usual to consider WIMPs distributed in a spherical 
halo around our galaxy with a local density $\rho = 0.3$ GeV/cm$^{3}$, a Maxwellian velocity distribution with
v$_{rms} \approx$ 270 km/s, an escape velocity of 650 km/s and the velocity of the Sun in the halo of 
230 km/s.

\subsection{Direct detection principle}
\label{subsec:direct}

In this technique, a WIMP is detected by measuring the nuclear recoil produced by its elastic interaction in 
an ordinary matter target.\\
In the MSSM model, the WIMP can couple to nuclei via 2 mechanisms~\cite{halo}:
\begin{itemize}
  \item spin-dependent, with $\sigma_{SD} \propto J(J+1)$, where J is the target
  nuclear spin
  \item spin-independent, with $\sigma_{SI} \propto A^2$, where A is the target
  atomic mass
  \end{itemize}  
In most MSSM models, $\sigma_{SI}$ dominates over $\sigma_{SD}$~\cite{halo} for A$>$30 (as it is for most of the 
used targets).\\
The most striking signature of WIMP interaction is the fact that it induces a nuclear recoil, while the 
natural radioactive background (except neutrons) leads to electronic recoils.\\ 
With sufficient statistics, other signatures can be used. The WIMP recoil energy spectrum 
should have the expected exponential behavior. In addition, the nuclear recoil directions should be 
correlated with the 
motion of the Sun in the galaxy. Although the measurement of recoil directions is difficult~\cite{zeplin}, it
could be possible to observe an annual modulation effect on the WIMP rate, as the Earth revolves around the 
Sun.
To remove nuclear recoils due to neutron collisions, two following signatures can be used : because of the 
weak interaction probability in matter, a WIMP can't produce multiple scattering (contrary to a neutron), and 
the spin-independent WIMP-nucleon cross section is proportionnal to $A^2$, while the dependence for neutron 
is approximately $A^{2/3}$.\\
When a WIMP hits a nucleus, its recoil energy is given by :
\begin{equation}
\label{eq:ER}
E_R = \frac{\mu^2}{M} v^2 (1-\cos\theta)
\end{equation}
where $M$ is the nucleus mass, $\mu$ the reduced mass ($M,m$) with $m$ the WIMP mass, $v$ the WIMP velocity and 
$\theta$ the scattering angle of WIMP. Given the velocity of WIMPs in our halo ($\sim$ 10$^{-3}$c) and the 
interesting mass range (GeV/c$^2$ to TeV/c$^2$), the energy  deposited by the particle is very low (10-100 
keV). In addition, supersymmetric calculations predict that the interaction rate should be very low (from 1 
evt/day/kg of detector to 1 evt/decade/kg).\\
These two facts lead to some constraints for the detectors. The low interaction rate requires a large 
detector mass and a very low background while the low deposited energy 
requires a low detection threshold ($\sim$ 10 keV recoil).\\
When a particle hits matter, according to the nature of the target, three different physical effects can be 
measured : ionization, scintillation and heat. Some experiments measure only one process and others measure 
two 
processes, in order to discriminate electronic and nuclear recoils (see Table \ref{tab1}).
 This discrimination is based on the fact 
that nuclear recoils produce proportionnally less ionization and scintillation than electronic recoils.
Because of these differences in responses, in this paper two kinds of
energy are quoted : the recoil energy in keV and the ionization or scintillation energy in \kevee (keV
electron equivalent); for Ge $ \frac{keV}{keV_{ee}} \approx$ 0.33 and for I $ \frac{keV}{keV_{ee}} \approx$
0.09.\\
Of these different techniques, we will present Germanium diodes, scintillating detectors and cryogenic 
detectors \footnote{We present here only the experiments which have published a limit and are 
sensitive to the spin-independent cross section}.
\begin{table}[h]
\caption{\label{tab1} Running experiments}
\begin{center}
\begin{tabular}{|c|c|c|c|}
\hline
Experiment & Location & Technique & Detectors\\
\hline
EDELWEISS I & Laboratoire Souterrain & ionization-heat & 3 $\times$ 320 g of Ge\\
 & de Modane & & \\
\hline
CDMS I & Standford & ionization-heat & 4 $\times$ 165 g of Ge \\
 & Underground Facility & & 1 $\times$ 100 g of Si\\
 \hline  
ZEPLIN I & Boulby mine & scintillation & 4 kg of LiqXe\\
\hline
CRESST I & Gran Sasso & scintillation-heat & 6 g of CaWO${_4}$\\
\hline
IGEX & Canfranc & ionization & 2 kg of Ge \\
\hline
HDMS & Gran Sasso & ionization & 200 g of Ge \\
\hline
DAMA & Gran Sasso & scintillation & 100 kg of NaI \\
\hline
\end{tabular}
\end{center}
\end{table} 

\subsection{Classical Germanium detector}
\label{subsec:Ge}

These detectors are Germanium diodes with two electrodes for collecting the charge induced by a particle 
interaction. The first search made with these detectors was that of neutrinoless 
double-beta decay ($0\nu 2\beta$). Their advantages are their energy resolutions and the lowest total
background event 
rate of any Dark Matter search thanks to improvements in the Germanium purification techniques. \\
The most important experiments are HDMS~\cite{hdms} and IGEX~\cite{igex}.\\
The HDMS (Heidelberg Dark Matter Search) detector is devoted to WIMP detection.
It is located at the Gran Sasso Underground Laboratory (LNGS) and comprises a 200 g natural Germanium detector 
surrounded by a 2.1 kg 
Germanium veto detector. With a resolution of 1.06 keV$_{ee}$ at 0 keV$_{ee}$ (determined by 
extrapolation) and an energy threshold of 2 keV$_{ee}$ for the inner detector, they recorded a background of 
0.2 evt/keV$_{ee}$/kg/d~\cite{hdms} in the range 11-40 keV$_{ee}$. The limit deduced from these data is not yet 
competitive with their previous limit~\cite{hdms3} : 0.042 evt/keV$_{ee}$/kg/d in the range 15-40 keV$_{ee}$ 
(with an energy thresholf of 9 keV$_{ee}$) obtained with a Germanium crystal enriched in 
$^{76}$Ge. Another stage was reached in 2002 with the first result in the final setup~\cite{hdms2} where 
 the inner detector was replaced by enriched $^{73}$Ge. They have obtained 0.43 evt/keV$_{ee}$/kg/d in the
range 11-40 keV$_{ee}$ (with an energy thresholf of 4 keV$_{ee}$ for the inner detector).\\
The IGEX detector is a 2.2 kg Germanium crystal enriched in $^{76}$Ge for the $0\nu 2\beta$
decay search, operated at the Canfranc Underground Laboratory. Their latest results 
correspond to an exposure of 80 kg.day~\cite{igex}. With an energy threshold of 4 keV$_{ee}$ and a 
resolution of 800 eV$_{ee}$ at 75 keV$_{ee}$, they obtained a background of $\sim$ 0.06 evt/keV$_{ee}$/kg/day 
\cite{igex} between 10-40 keV$_{ee}$. Their data exclude a WIMP with a mass of $\sim$ 50 GeV/c$^2$ and a 
WIMP-nucleon cross section $\sigma_{\chi -n}$ $\approx$ 7 $\times$ 10$^{-6}$ pb. For experiments using Ge
diodes, this is the highest Dark Matter sensitivity.\\
The disadvantage of this Germanium technique for Dark Matter direct detection is the impossibility to 
discriminate nuclear and electronic recoils.   

\subsection{Scintillating detectors}
\label{subsec:scint}

The second category of direct detection detectors includes liquid and solid scintillators. The 
running experiments use detectors with already large masses (5 to 100kg).\\
The DAMA detector~\cite{dama} is made with 100 kg of NaI crystals placed at the Gran Sasso Underground
Laboratory (LNGS). In four years, they accumulated an exposure of 58000 kg.day~\cite{dama}. Their statistical
background rejection is based on annual modulation. In 2000 this team claimed to observe an 
annual modulation~\cite{dama} of their event rate, in the 2-6 \kevee range, corresponding 
to a WIMP 
with $m$ = 52 GeV/c$^2$ and $\sigma_{\chi -n}$ = 7.2 $\times$ 10$^{-6}$ pb using standard astrophysical 
assumptions.
\begin{figure}[ht]
\begin{center}
\includegraphics*[width=12cm,height=6cm]{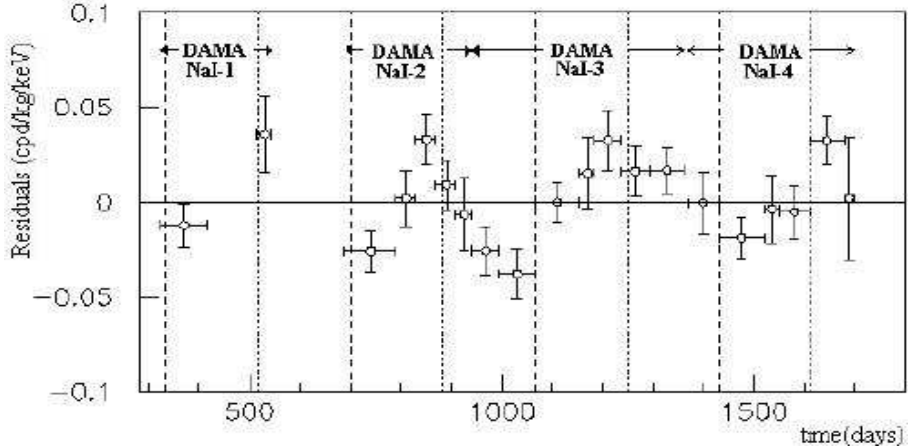}
\end{center}
\caption{\label{rot}{Annual modulation of the count rate as observed by DAMA~\protect\cite{dama} in
four years}}
\end{figure}
In combining this result with their previous exclusion curve of 1996~\cite{dama2}, their most likely WIMP
mass is 44 GeV/c$^2$ with $\sigma_{\chi -n}$ = 5.4 $\times$ 10$^{-6}$ pb.\\
The ZEPLIN (ZonEd Proportionnal scintillation in LIquid Noble gases) collaboration uses another type 
of scintillator : the liquid Xenon. This detector is located at the Boulby mine. The first stage, ZEPLIN I, 
consists in a liquid Xe single phase with a fiducial mass of $\sim$ 4 kg. For more details on its 
preliminary results, see~\cite{zeplin}. A second stage~\cite{zeplin2} is in preparation with 2 phases 
(liquid-gas) allowing the simultaneous measurement of scintillation and ionization with a Xe target 
of $\sim$ 30 kg. 

\subsection{Cryogenic detectors}
\label{subsec:cryo}

When a particle hits a crystal (the absorber), its temperature increases by $\Delta T = \frac{\Delta E}{C}$,
where C is the heat capacity of the crystal and $\Delta E$ is the deposited energy. To measure 
this tiny increase, the detector has to be placed at a cryogenic temperature of $\sim$ 10 mK in order to 
minimize C. $\Delta$T is measured by a sensor glued, or evaporated, on the absorber. By measuring the heat signal in 
coincidence with another signal such as ionization or scintillation, it is
possible to discriminate nuclear and electronic recoils, and thus to remove a major part of the
background.

\subsubsection{The CRESST experiment}
\label{cresst}

The CRESST detector is located at the Gran Sasso Underground Laboratory (LNGS). It measures at the same time the 
scintillation and the heat signals produced by the incident particle to perform a discrimination. Their detectors 
have three parts : a scintillating absorber made of Calcium Tungstate, a Sapphire light detector and a 
Tungsten thermometer placed on the absorber. The latest results~\cite{cresst} of the collaboration were 
obtained with a 262 g Sapphire crystal with a Tunsgten film as sensor, operating at 15 mK. At present, two 
detectors each with a mass of 300 g are running. 

\subsubsection{Heat and ionization bolometers}
\label{bolo}

\begin{figure}[ht]
\begin{center}
\includegraphics*[width=7cm,height=7cm]{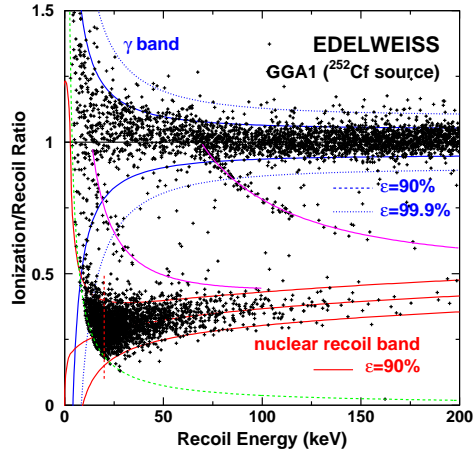}
\end{center}
\caption{\label{calib}{Plot of the ratio of the ionization energy to the recoil 
energy, versus the recoil energy recorded in an EDELWEISS~\protect\cite{edelweiss2} Ge detector using a
${^{252}}$Cf source. The full lines represent 90$\%$ of the 2 populations (nuclear and electron recoils).}}
\end{figure}

The CDMS and EDELWEISS experiments use bolometers with simultaneous measurements of ionization and heat. \\
The interaction of a particle in the semiconducting absorber (Silicium or Germanium crystal) produces 
ionization collected by two electrodes. The rise in temperature is measured with a heat sensor (NTD (Neutron
Transmutation Doped) or TES (Transition Edge Sensor)). These two simultaneous measurements 
provide an event-by-event discrimination between nuclear and electronic recoil.\\ 
The principle of this discrimination is as follows. The ratio of the ionization and heat signals 
depends on the recoiling particle, since a nucleus produces less ionization in crystal than an electron. 
On a plot of the ratio of the ionization energy to the recoil energy, versus the recoil energy, two 
populations can be distinguished (Fig. \ref{calib}). 
The first population centered on $\sim$ 1 (by construction) represents electron recoils from gamma 
interactions, while the second one centered on $\sim$ 0.3 corresponds to nuclear recoils, in this case from 
neutron scattering. 
Fig. \ref{calib} shows an example where it is possible to reject more than $\sim$ 99.9 $\%$ of gammas 
down to 15 keV recoil energy.   

\subsubsection{The CDMS experiment}
\label{cdms}

Until last year the CDMS~\cite{cdms} experiment was located at the Stanford Underground Facility. Their setup
includes two different crystal absorbers (4 $\times$ 165 g Germanium and 1 $\times$ 100 g Silicium).
In 1998 and 1999 CDMS performed some runs with these Germanium bolometers (Fig. \ref{data}) with an analysis 
threshold of 10 keV. They recorded 11.9 kg.day (in the inner volume) after subtracting neutron 
background on the basis of the relative rate in Si and Ge, and the coincident rate of nuclear recoils between 
Ge detectors. Their data exclude at more than 99 $\%$ C.L. the DAMA experiment candidate with $m$ = 
52 GeV/c$^2$ and $\sigma_{\chi -n}$ = 7.2 $\times$ 10$^{-6}$ pb, and moreover, exclude a large part of the 
DAMA region (Fig. \ref{exclu}). \\
In 2002 CDMS moved to the Soudan mine for a better protection against neutrons induced by cosmic rays. 
Data taking is in progress.
 
\section{The EDELWEISS experiment}
\label{edelweiss}

The EDELWEISS collaboration combines several laboratories from CEA (DAPNIA, DRECAM), CNRS (IN2P3, DSM, INSU) 
and Germany (FZKA, Karlsruhe). EDELWEISS means "Exp\'erience pour DEtecter Les Wimps En SIte Souterrain" and is located at the LSM 
(Laboratoire Souterrain de Modane) in the Frejus tunnel under the Franco-Italian Alps. Their latest results 
were published in 2002~\cite{edelweiss2}.

\subsection{The experimental setup}
\label{setup}  

The present stage of the EDELWEISS experiment~\cite{edelweiss2} consists of three 320 g Germanium cryogenic 
detectors placed in a dilution cryostat with a base temperature of $\sim$ 17 mK. To decrease the background 
in the experiment, all materials around the detectors were carefully selected for their low radioactivity. 
The front end electronic components are placed behind a roman lead shield above the three detectors.
The Germanium bolometers are equipped with a NTD heat sensor. The ionization is 
collected by two Aluminium sputtered electrodes operated at voltages between 2 and 4 V. One of the three 
detectors has a Germanium amorphous layer under the electrodes.\\
A segmented electrode defines a central fiducial part and a guard ring. Most of the radioactivity due to the 
detector environment is collected on this latter part of the detector. In addition, the electrostatic field 
is more uniform in the central part. The fiducial inner volume, defined as $\ge$ 75 $\%$ of the charge 
collected on the central electrode, corresponds to 57$\%$ of the total detector volume~\cite{olivier}.  

\subsection{Results}
\label{results}  

Calibrations with gamma-ray sources revealed that two of the three detectors had problems with the charge 
collection, that were related with the absence of amorphous layer. Therefore only the third detector data 
were used for the final results.\\
The low-background physics runs between February and May 2002 resulted in 8.6 kg.day effective exposure in 
the fiducial volume of the selected detector. The baseline resolutions for this detector are below 1.5 \kevee 
and 1.3 \kevee for ionization and heat respectively. The ionization threshold  corresponding to an efficiency 
of 50$\%$ is measured to be 3.7 $\pm$ 0.2 \kevee, implying a full efficiency at a recoil threshold of 20 keV.
\\ 
No events have been observed in the band corresponding to 90$\%$ of all nuclear recoils (Fig. \ref{data}). 
One event lies on the edge of the nuclear recoil zone at recoil energy $\sim$ 120 keV but is incompatible 
at 95$\%$ C.L. with a WIMP with a mass below 10 TeV/c$^2$. These data exclude at more than 99.994 
$\%$ the DAMA candidate with $m$ = 52 GeV/c$^2$ and $\sigma_{\chi -n}$ = 7.2 $\times$ 10$^{-6}$ pb.
  
\subsection{Comparison with the CDMS results}
\label{comparison}

\begin{figure}[ht]
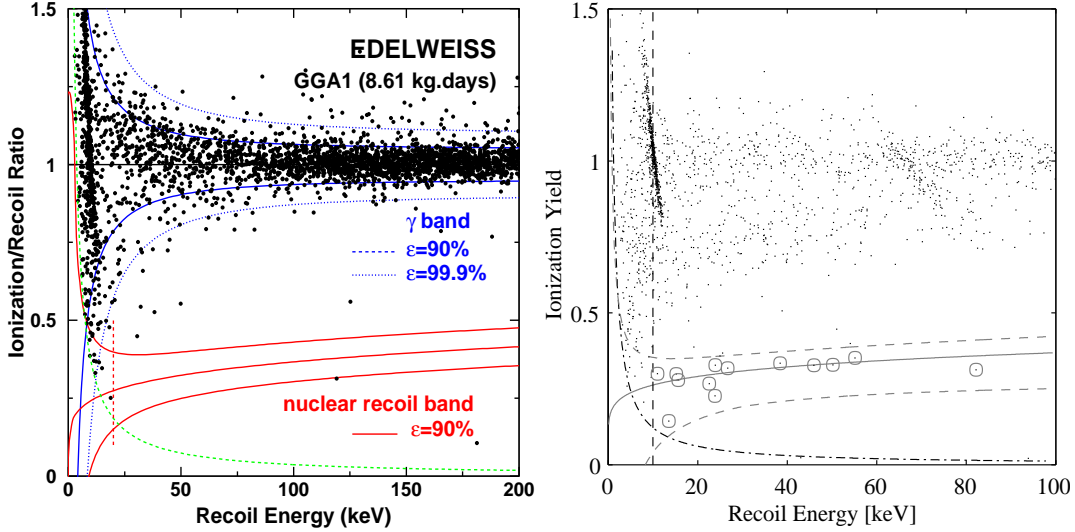

\begin{center}
\mbox{
\includegraphics*[width=7cm,height=7cm]{fond2002.epsi}
\includegraphics*[width=7cm,height=7cm]{cdms1.epsi}}
\end{center}
\caption{\label{data}{Distribution of the ratio of the ionization to the recoil
energy as a function of the recoil energy for the EDELWEISS~\protect\cite{edelweiss2} (8.6 kg days) and
CDMS~\protect\cite{cdms} (11.9 kg days) experiments.}}
\end{figure}

EDELWEISS and CDMS using the same type of detectors, it is interesting to compare their results 
(Fig. \ref{data}).
EDELWEISS observed no events in the nuclear recoil band, in the range 20-200 keV, in 8.6 
kg.day~\cite{edelweiss2} while CDMS observed 13 events, in the range 10-100 keV, in 11.9 kg.day~\cite{cdms}. 
CDMS attributes these events to a 
neutron background based on their Silicium detectors data (compared to Ge, Si is relatively more sensitive to 
neutrons than to WIMPs) and the coincident rate between Germanium detectors. This background arises because 
of their shallow site. EDELWEISS is located under 1600 meters of rock, which reduces the prompt activation 
by cosmic rays by a factor one million.
 
\subsection{Exclusion limit}
\label{exclusion}

\begin{figure}[ht]
\begin{center}
\includegraphics*[width=7cm,height=7cm]{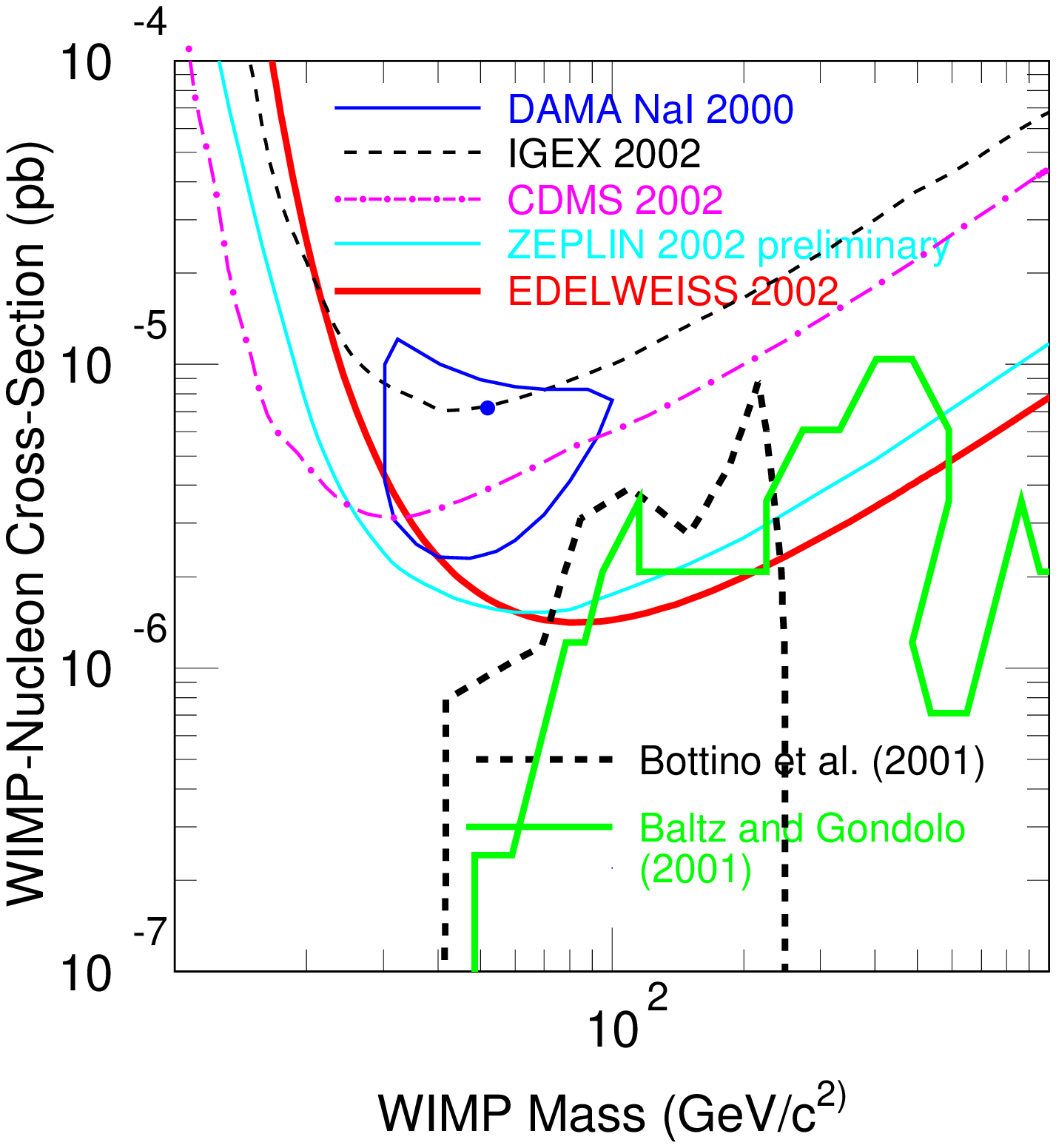}
\end{center}
\caption{\label{exclu}{90$\%$ C.L. exclusion limit for the experiments DAMA~\protect\cite{dama}, 
IGEX~\protect\cite{igex}, CDMS~\protect\cite{cdms}, EDELWEISS~\protect\cite{edelweiss2} and 
ZEPLIN~\protect\cite{zeplin} (preliminary) and two regions spanned by some supersymmetric 
calculations~\protect\cite{{bottino},{baltz}} are shown}}
\end{figure}

Finally all experiments compare the observed rate to the predicted one and interpret the results as 
90$\%$ 
C.L. exclusion limits on the WIMP-nucleon cross section as a function of the WIMP mass (Fig. \ref{exclu}). 
To be able to compare various experiments, all of them assume the same parameters for the halo model.
The EDELWEISS experiment exludes at 99.8$\%$ the DAMA candidate with $m$ = 44 GeV/c$^2$ and 
$\sigma_{\chi -n}$ = 5.4 $\times$ 10$^{-6}$ pb. Its sensitivity for spin-independent WIMP-nucleon interaction 
is the best published result for masses above 35 GeV/c$^2$. Moreover, their curve start to covers some 
supersymmetric predictions. Experiments with event-by-event discrimination are today more sensitive than 
experiments with classical detectors.\\
Note that the ZEPLIN curve is still preliminary. 

\section{Outlook}
\label{futur}

\subsection{Coming developments}

All the described experiments are preparing for the near future new stages with increased masses and improved 
techniques (see Table \ref{tab2}). \\
For example, the EDELWEISS experiment is running in 2002-03 with three 320 g Germanium bolometers, all of 
them having an amorphous layer (either in Germanium or Silicium).
\begin{table}[h]
\caption{\label{tab2} Future experiments} 
\begin{center}
\begin{tabular}{|c|c|c|c|}
\hline
Experiment & Location & Technique & Detectors\\
\hline
EDELWEISS II & Laboratoire Souterrain & ionization-heat & 120 $\times$ 320 g of Ge\\
 & de Modane & & \\
\hline
CDMS II & Soudan mine & ionization-heat & 21 $\times$ 250 g of Ge \\
 & & & 21 $\times$ 100 g of Si\\
\hline  
ZEPLIN II & Boulby mine & scintillation-ionization & 30 kg of LiqXe\\
\hline
CRESST II & Gran Sasso & scintillation-heat & 33 $\times$ 300 g of CaWO${_4}$\\
\hline
GENINO & Gran Sasso & ionization & 100 kg of Ge \\
\hline
LIBRA & Gran Sasso & scintillation & 250 kg of NaI \\
\hline
\end{tabular}
\end{center} 
\end{table}

\subsection{EDELWEISS II}

EDELWEISS II will be an experiment with 120 (40 kg) Germanium cryogenic detectors placed in a new cryostat. 
The detectors will be placed in an innovative reversed cryostat with a volume of 100 $\ell$. Inside, the 
detectors are close packed in an hexagonal arrangement. The cryostat has been already tested at $\sim$10 mK. 
At present, a first phase with 28 detectors is approved\\
The goal of this experiment is to improve by two orders of magnitude the present sensitivity. In order to
reduce the neutron background below $\sim$ 0.02 evt/kg/day, the shielding will be improved by
surrounding the experiment with muon vetos and by increasing the thickness of the Polyethylene shield. 

\subsection{Sensitivity goals}

\begin{figure}[ht]
\begin{center}
\includegraphics*[width=7cm,height=7cm]{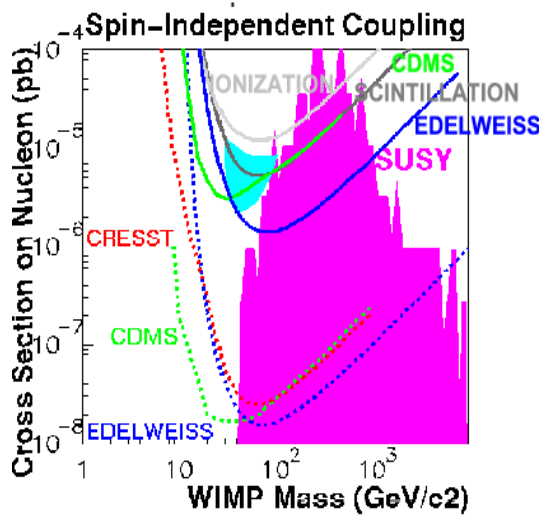}
\end{center}
\caption{\label{limit}{Spin independent sensitivities expected from next generation hybrid calorimetric
experiments (figure adapted from ~\protect\cite{edelweiss3}).}}
\end{figure}

Dark Matter direct detection experiments are now reaching WIMP-nucleon cross sections of $10^{-6}$ pb and 
start to probe the most optimistic supersymmetric models. But testing a more significant fraction of the 
models requires to reach a WIMP-nucleon cross section of $10^{-8}$ pb at least (Fig. \ref{limit}). This is 
the main purpose of the second phases of the experiments. 
In this aim, they increase the target masses and work to improve their sensitivity with new materials and new 
techniques. 
 
\section{Conclusion}

At present, WIMP direct detection experiments start now to be sensitive to a small fraction of the most 
optimistic supersymmetric models. Next 
generation of experiments should allow a factor 100 improvement in sensitivity and begin to test a larger
fraction of models. But testing the bulk of supersymmetric models requires experiments in the one-ton 
range and an extreme background rejection. All the experiments aim at discovering the nature of the main 
part of Dark Matter in the coming years, to possibly confirm a signal observed with the LHC after 2007.


\end{document}